\documentclass[english,american,aps,groupedaddress,prl,preprint]{revtex4-1}
\usepackage[T1]{fontenc}
\usepackage[latin9]{inputenc}
\usepackage{amstext}
\usepackage{graphicx}
\usepackage{esint}

\makeatletter
 
 \@ifundefined{textcolor}{}
 {%
   \definecolor{BLACK}{gray}{0}
   \definecolor{WHITE}{gray}{1}
   \definecolor{RED}{rgb}{1,0,0}
   \definecolor{GREEN}{rgb}{0,1,0}
   \definecolor{BLUE}{rgb}{0,0,1}
   \definecolor{CYAN}{cmyk}{1,0,0,0}
   \definecolor{MAGENTA}{cmyk}{0,1,0,0}
   \definecolor{YELLOW}{cmyk}{0,0,1,0}
 }

\@ifundefined{definecolor}
 {\usepackage{color}}{}

\makeatother

\usepackage{babel}
\begin{document}

\title{Characterization of predator-prey dynamics, using the evolution of
free energy in plasma turbulence}

\author{P. Morel}

\affiliation{Laboratoire de Physique des Plasmas, Ecole Polytechnique, CNRS, 91128
Palaiseau Cedex, France.}

\author{\"O. D. G\"urcan}

\author{V. Berionni}

\affiliation{Laboratoire de Physique des Plasmas, Ecole Polytechnique, CNRS, 91128
Palaiseau Cedex, France.}

\begin{abstract}
A simple dynamical cascade model for the evolution of free energy
is considered in the context of gyrokinetic formalism. It is noted
that the dynamics of free energy, that characterize plasma micro-turbulence
in magnetic fusion devices, exhibit a clear predator prey character.
Various key features of predatory prey dynamics such as the time delay
between turbulence and large scale flow structures, or the intermittency
of the dynamics are identifed in the quasi-steady state phase of the
nonlinear gyrokinetic simulations. A novel prediction on the ratio
of turbulence amplitudes in different parts of the wave-number domain
that follows from this simple predator prey model is compared to a
set of nonlinear simulation results and is observed to hold quite
well in a large range of physical parameters. Detailed validation
of the predator prey hypothesis using nonlinear gyrokinetics provides
a very important input for the effort to apprehend plasma microturbulence,
since the predator prey idea can be used as a very effective intuitive
tool for understanding.
\end{abstract}

\maketitle

Predator-prey interactions are a very common paradigm in natural sciences,
which provide a powerful perspective for the interpretation of various
complex phenomena\citep{goel:1971}. In the context of fusion plasmas,
the evolution of turbulence and self-regulating sheared flows that
it drives, may eventually explain the dynamical coupling leading to
the Low to High confinement (L-H) transition \citep{malkov:01,kim:03}
in magnetic fusion devices. Interesting quasi-periodic activity, which
may be linked to the predator-prey oscillations between turbulence
and large scale flows in the form of zonal flows or geodesic acoustic
modes, (GAMs) has been observed recently in a number of machines prior
to and during the L-H transition\citep{estrada:10,conway:11,Zhao:10}.
The predator-prey dynamics also plays an important role in the nonlinear
cascade process via the refraction of the turbulence in the low-$k$
($k$ being the wave-number) energy containing scales to high-$k$
dissipative scales by the self-generated zonal flows \citep{smolyakov:00,gurcan:09,berionni:11}.
This mediating role of zonal flows during the cascade of free energy
has also recently been observed in gyrokinetic simulations \citep{nakata:12}.

Plasma turbulence in a strong magnetic field can be described by the
gyrokinetic equation \citep{catto:78,lee:83,hahm:88}, which by filtering
the rapid gyromotion, reduces the Vlasov equation to resolving a five-dimensional
distribution function $f=f({\bf R},v_{\parallel},\mu,t)$, where ${\bf R}$
is the guiding center coordinate, $v_{\parallel}$ is the velocity
coordinate along the magnetic field $B_{0}$ and $\mu=m_{i}v_{\perp}^{2}/(2B_{0})$
is the magnetic moment, which is an adiabatic invariant. In a field
aligned geometry (${\bf R}\rightarrow x,y,z$ with $z$ the field
aligned, $x$ the radial and $y$ the binormal coordinate), and separating
between fluctuations and a Maxwellian equilibrium $f=F_{0}+\delta f$,
the gyrokinetic system of equations read:\vspace{-0.5cm}

\begin{eqnarray}
\partial_{t}\delta f_{k} & = & i\omega_{\star Ti}F_{0}J_{0}\Phi_{k}+i\omega_{D}\delta h_{k}-D[\delta h_{k}]\nonumber \\
 & + & v_{Ti}\left(\frac{\mu\partial_{z}B_{0}}{2}\partial_{v_{\parallel}}\delta h_{k}-v_{\parallel}\partial_{z}\delta h_{k}\right)\nonumber \\
 & + & \sum_{{\bf p}+{\bf q}=-{\bf k}}\left(\hat{{\bf b}}\text{\ensuremath{\times}{\bf p}\,}.\,{\bf q}\right)J_{0}\Phi_{{\bf p}}^{\star}\delta h_{{\bf p}}^{\star}\label{eq:GKt}
\end{eqnarray}
\begin{equation}
\int d\mu dv_{\parallel}J_{0}\delta h_{k}=\widetilde{\Phi}_{k}-\frac{T_{e0}}{T_{i0}}\Phi_{k}\,,\label{eq:QN}
\end{equation}
where $\delta h_{k}=\delta f_{k}+\frac{q_{i}F_{0}}{T_{0i}}J_{0}\Phi_{k}$
has been used for compactness, the unknowns $\delta f_{k}$ and $\Phi_{k}$
are Fourier transformed in the plane $(k_{x},k_{y})$. Here, $\widetilde{\Phi}_{k}=\Phi_{k}-\left\langle \Phi_{k}\right\rangle $,
where $\left\langle \Phi_{k}\right\rangle $ is the flux surface averaged
electrostatic potential. $J_{0}=J_{0}(k_{\perp}\rho_{\perp})$ is
the Bessel function of order zero\textcompwordmark{}, where $k_{\perp}(z)$
is the perpendicular wave number, and $\rho_{\perp}(z,\mu)=v_{Ti}\frac{\sqrt{B_{0}(z)\mu}}{\Omega_{ci}(z)}$.
Electrons are assumed adiabatic with temperature $T_{e0}$ and ions
are characterized by an equilibrium density $n_{i0}$ and a temperature
$T_{i0}$ (with $T_{i0}=T_{e0}$ in the following), $v_{Ti}=\sqrt{\frac{2T_{i0}}{m_{i}}}$,
$\Omega_{ci}=\frac{q_{i}B_{0}}{m_{i}}$ and $\rho_{i}=\frac{v_{Ti}}{\Omega_{ci}}$
are respectively ion thermal velocity, cyclotron frequency and Larmor
radius. Ion temperature and density equilibrium profiles are contained
in $\omega_{*Ti}=-\frac{k_{y}}{L_{n}}\left(1+\left(v_{\parallel}^{2}+\mu B_{0}-\frac{3}{2}\right)\frac{L_{n}}{L_{Ti}}\right)$,
where: $\frac{1}{L_{n}}=-\frac{dn_{i0}}{n_{i0}dr}$ and $\frac{1}{L_{Ti}}=-\frac{dT_{i0}}{T_{i0}dr}$
. The drift frequency $\omega_{D}=-\frac{v_{\parallel}^{2}+\mu B_{0}/2}{\Omega_{ci}/v_{Ti}^{2}}\left(K_{x}k_{x}+K_{y}k_{y}\right)$,
contains magnetic unhomogeneity, details about the magnetic geometry
can be found in Ref. \citep{lapillonne:09}. $D[\delta h_{k}]=c_{z}\partial_{z}^{4}\delta h_{k}+c_{v_{\parallel}}\partial_{v_{\parallel}}^{4}\delta h_{k}+c_{\perp}k_{\perp}^{2n}\delta h_{k}$
corresponds to numerical dissipations. Simulations that we present
in the following are performed using the GENE code\citep{jenko:00,dannert:05}.
Despite the fact that GENE is also adapted for electromagnetic and
global problems \citep{gorler:11}, the direct numerical simulations
(DNS) presented here are restricted to local electrostatic ion temperature
gradient driven turbulence (ITG) with adiabatic electrons. 

The gyrokinetic equation, as written in (\ref{eq:GKt}), has a number
of nonlinearly conserved quantities, one of which is the so called
free energy, which budget can be written as:\vspace{-0.5cm}

\begin{equation}
\partial_{t}\mathcal{E}=\mathcal{G}-\mathcal{D}\,,\label{eq:FE}
\end{equation}
where $\mathcal{E}=n_{i0}\int d\Lambda_{k}\frac{\delta h_{k}^{\star}\delta f_{k}}{2F_{0}}$,
$\mathcal{G}=n_{i0}\int d\Lambda_{k}i\omega_{*Ti}J_{0}\Phi_{k}\delta h_{k}^{\star}$,
and $\mathcal{D}=n_{i0}\int d\Lambda_{k}\frac{\delta h_{k}^{\star}}{F_{0}}D[\delta h_{k}]$
 respectively define the free energy, its injection and dissipation
(using the phase space integration $\int d\Lambda_{k}=\sum_{k_{x},k_{y}}\int\frac{\pi}{V}dzdv_{\parallel}d\mu$
with the volume $V=\sum_{k_{x},k_{y}}\int dz/B_{0}$). A local free
energy balance in perpendicular Fourier space can be expressed as:\vspace{-0.5cm}

\begin{equation}
\partial_{t}\mathcal{E}_{\ell_{\perp}}=\mathcal{G}_{\ell_{\perp}}+\mathcal{N}_{\ell_{\perp}}-\mathcal{D}_{\ell_{\perp}}\label{eq:FEkperp}
\end{equation}
where $\ell_{\perp}$ may be taken to correspond to any partition
of the perpendicular Fourier space (for example $k_{\perp\ell}<\ell_{\perp}<k_{\perp\ell+1}$),
and the contribution of the nonlinear term satisfies $\sum_{\ell_{\perp}}\mathcal{N}_{\ell_{\perp}}=0$.

In plasma turbulence, zonal flows\citep{diamond:05} are of special
importance, since these structures, extended over a given flux surface,
play a regulating role on the turbulence that generates them. The
zonal flow free energy $\overline{\mathcal{E}}$ can be separated
from the rest of the drift wave turbulence $\tilde{\mathcal{E}}$,
where the energy budget takes the form:\vspace{-0.5cm}

\begin{eqnarray}
\partial_{t}\overline{\mathcal{E}} & = & \overline{\mathcal{N}}-\overline{\mathcal{D}}\,,\label{eq:FE-ZF}\\
\partial_{t}\widetilde{\mathcal{E}} & = & \widetilde{\mathcal{G}}+\widetilde{\mathcal{N}}-\widetilde{\mathcal{D}}\,.\label{eq:FE-NZ}
\end{eqnarray}

\begin{figure}
\includegraphics[width=1\linewidth]{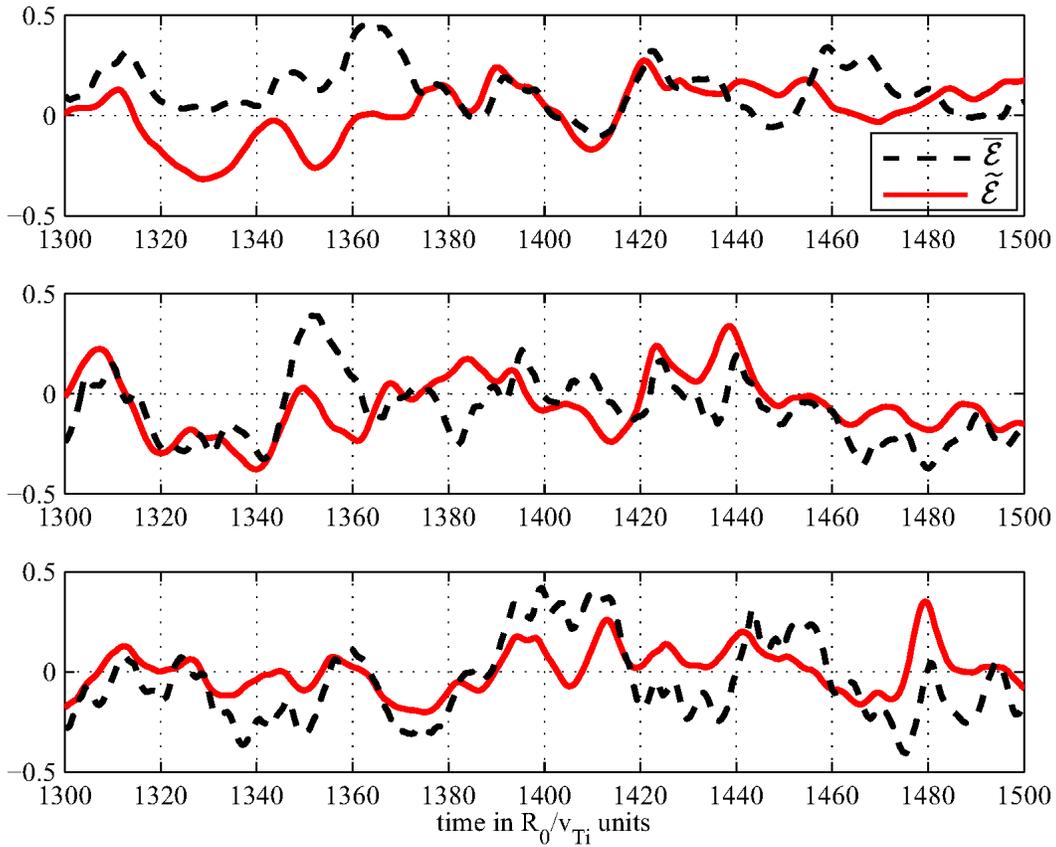}
\caption{\label{fig:ZF-NZ-time}Time evolution of the normalized free energy
$\overline{\mathcal{E}}$ associated with zonal flows and $\widetilde{\mathcal{E}}$
associated with the rest of the turbulence, for three different values
of $R_{0}/L_{Ti}=6.0,6.92,8.0$ from top to bottom.}
\end{figure}

\begin{figure}
\includegraphics[width=1\linewidth]{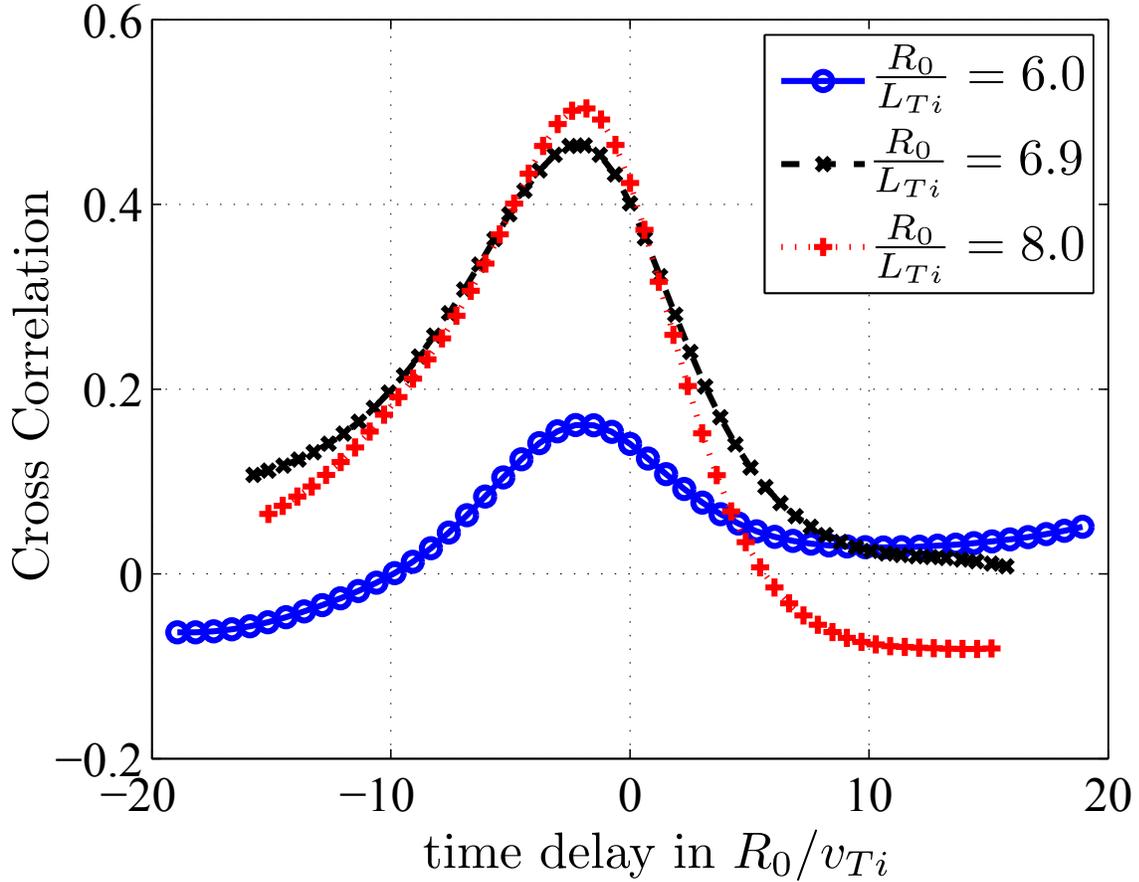}
\caption{\label{fig:ZF_NF_cross-corr}Cross Correlation between Zonal Flow
and drift wave turbulent signals, for different values of the logarithmic
temperature gradient $R_{0}/L_{Ti}$.}
\end{figure}

It is important to note here that it is the free energy (which corresponds
to ``potential enstrophy'' in the fluid limit) that is exchanged
between the zonal flows and the drift waves and not just the kinetic
energy. The mechanism invoked here is not that of a classical ``inverse
cascade'' but of a potential vorticity homogenization\citep{Rhines:1982,diamond:08b},
which manifests itself as disparate scale interactions in $k$-space\citep{krommes:00}.
Since there is no linear driving mechanism for the zonal flows (i.e.
$\overline{\mathcal{G}}=0$), these structures feed on the free energy
of the fluctuations and hence play the same regulating role on the
underlying turbulence that a predator species plays on the population
of a prey species. Figure \ref{fig:ZF-NZ-time} represents the time
evolution of free energy ($\overline{\mathcal{E}}$ and $\widetilde{\mathcal{E}}$
have been normalized to their mean, and only a small fraction of the
total time trace is represented in order to see the details of dynamics),
during the turbulent phase for three values of $R_{0}/L_{Ti}=6.0,6.92,8.0$,
with other parameters being those of the ITG Cyclone Base Case\citep{dimits:00}
($R_{0}/L_{n}=2.22$, $q=1.4$, $\hat{s}=0.796$, $r_{0}=0.18R_{0}$,
and $T_{e0}=T_{i0}$). It can be observed that the free energy dynamics
of the zonal flow and the turbulence are indeed largely correlated.
In order to check if this dynamics exhibit predator-prey features,
one can look at the phase relation between these two quantities to
see if there exists a time shift between the turbulence and the zonal
flow free energy fluctuations. In Figure \ref{fig:ZF_NF_cross-corr},
cross correlation between $\overline{\mathcal{E}}$ and $\widetilde{\mathcal{E}}$
is given as a function of the time lag. The average time delay $\Delta t$
between $\overline{\mathcal{E}}$ and $\widetilde{\mathcal{E}}$ is
given by the location of the maximal correlation in Figure \ref{fig:ZF_NF_cross-corr}.

\begin{figure}
\includegraphics[width=1\linewidth]{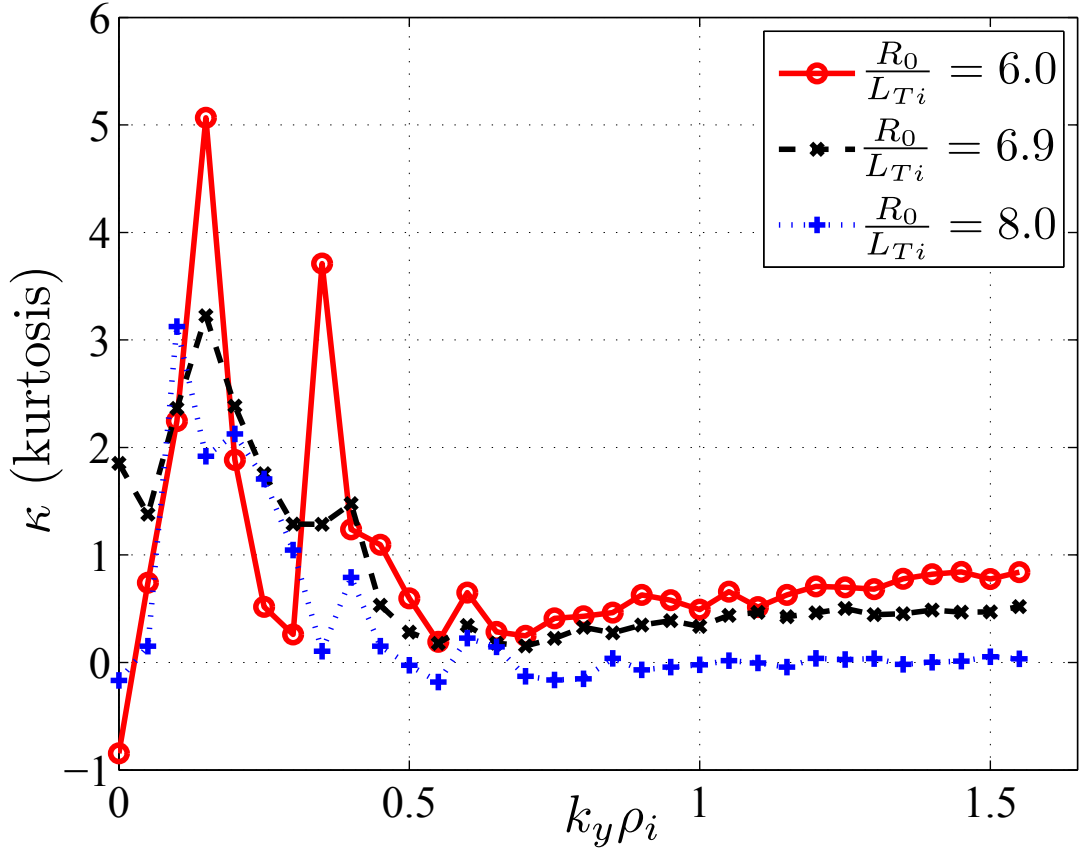}
\caption{\label{fig:FE-kurt-ky-LT} Free Energy kurtosis $\kappa$ as a function
of the binormal wave vector $k_{y}\rho_{i}$, for different values
of the logarithmic ion temperature gradient $R_{0}/L_{Ti}$.}
\end{figure}

The predator prey type dynamics is also expected to have an intermittent
nature. Therefore a look at the kurtosis is instructive: in Figure
\ref{fig:FE-kurt-ky-LT}, the kurtosis associated with the free energy
spectrum $\mathcal{E}^{k_{y}}$ is plotted as a function of $k_{y}$
for the same runs as in Figure \ref{fig:ZF-NZ-time}. For all values
of the temperature gradient, a clear separation is observed between
the large and small scales, where the statistics of the small scales
are very close to Gaussian (corresponding to a kurtosis of 0), while
the energy containing scales associated with wave vectors $k_{y}\rho_{i}<0.5$
show a clear departure from the Gaussian distribution, suggesting
the presence of rare events with significant deviation from the mean.

\begin{figure}
\includegraphics[width=1\linewidth]{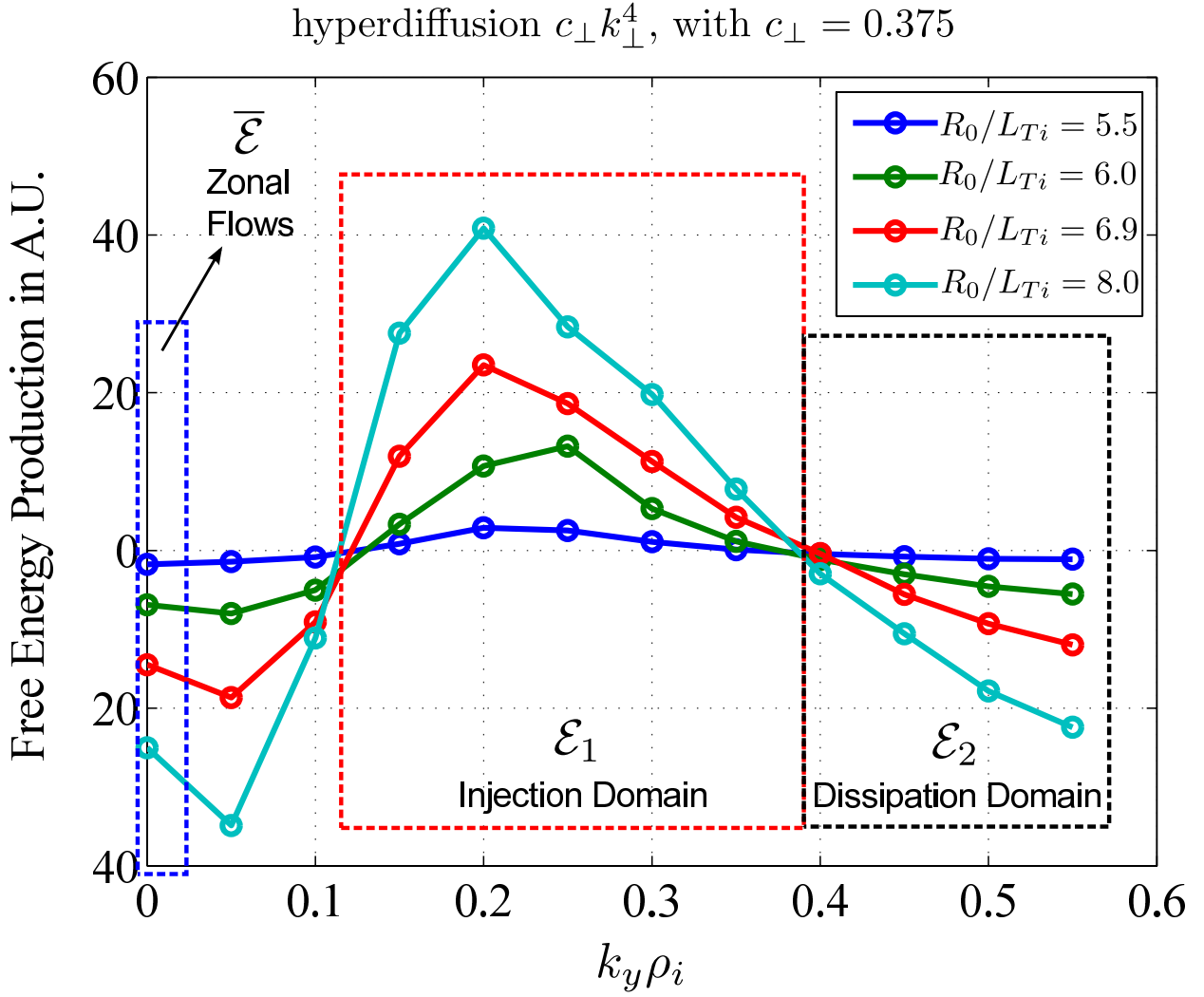}
\caption{\label{fig:FErhs-ky-LT} Free energy spectrum $\mathcal{E}^{k_{y}}$
for different values of the ion temperature gradient $R_{0}/L_{Ti}$.}
\end{figure}

While the usual predator prey model already gives interesting perspective
on the dynamics the fact that the predator-prey oscillations has the
character of pulses in the $k$-space transfer has to be considered
also. Indeed, the free energy can be decomposed into the ``energy
containing'' component (corresponding to the energy injection scales),
the ``dissipative'' component (corresponding to the dissipative
scales) and a ``Zonal Flow'' component. These three domains are
illustrated in Figure \ref{fig:FErhs-ky-LT}, where the time averaged
right hand side of the free energy equation i.e. $\mathcal{G}^{k_{y}}-\mathcal{D}^{k_{y}}$
is given for various values of the logarithmic temperature gradient
$R_{0}/L_{Ti}$ (the transfer term $\mathcal{T}^{k_{y}}$ is not represented
since: $\mathcal{T}^{k_{y}}=\mathcal{D}^{k_{y}}-\mathcal{G}^{k_{y}}$
\citep{navarro:11a}). Simulations have been performed on reduced
grids ($N_{x}\times N_{y}=48\times24$) by means of the GyroLES technique\citep{morel:11}
with a perpendicular hyperdiffusion model $c_{\perp}=0.375$.

The spectral transfer character of the predator-prey dynamics (\ref{eq:FE-ZF},
\ref{eq:FE-NZ}) can be studied by using a model of the form:\vspace{-0.5cm}

\begin{eqnarray}
\partial_{t}\overline{\mathcal{E}} & = & \overline{\mathcal{N}}-\nu_{F}\overline{\mathcal{E}}\,,\label{eq:FE-ZFbis}\\
\partial_{t}\mathcal{E}_{1} & = & \mathcal{N}_{1}+\gamma\mathcal{E}_{1}\,,\label{eq:FE-inj}\\
\partial_{t}\mathcal{E}_{2} & = & \mathcal{N}_{2}-\nu\mathcal{E}_{2}\,,\label{eq:FE-dis}
\end{eqnarray}
where, $\mathcal{E}_{1}$ is the free energy at the injection scale,
$\mathcal{E}_{2}$ is the free energy at the dissipation scale and
$\overline{\mathcal{E}}$ is the zonal free energy, and we have used
the definitions: $\nu_{F}=\overline{\mathcal{D}}/\overline{\mathcal{E}}$,
$\gamma=(\mathcal{G}_{1}-\mathcal{D}_{1})/\mathcal{E}_{1}$ and $\nu=(\mathcal{D}_{2}-\mathcal{G}_{2})/\mathcal{E}_{2}$.

This three domain model is very similar to the model studied in Ref.
\citealp{berionni:11}, except here we use free energy. As recently
shown by Nakata et al. the free energy contribution of the nonlinear
term can be expressed as a symmetrized triad transfer function: $\mathcal{N}_{{\bf k}}=\sum_{{\bf p}}\sum_{{\bf q}}C_{{\bf k}}^{{\bf p},{\bf q}}\delta h_{{\bf k}}\delta h_{{\bf p}}\delta h_{{\bf q}}$,
where $C_{{\bf k}}^{{\bf p},{\bf q}}$ is an operator converting the
modified distribution function $\delta h_{k}$ into the electrostatic
potential $\Phi_{k}$. This allows to write:\vspace{-0.5cm}

\begin{eqnarray}
\overline{\mathcal{N}} & = & \overline{\lambda}\,\overline{h}h_{1}h_{2}\,,\label{def:lambda-bar}\\
\mathcal{N}_{1} & = & \lambda_{1}\,\overline{h}h_{1}h_{2}\,,\label{def:lambda1}\\
\mathcal{N}_{2} & = & \lambda_{2}\,\overline{h}h_{1}h_{2}\,.\label{def:lambda2}
\end{eqnarray}
where $h_{1}$, $h_{2}$ and $\overline{h}$ can be defined for instance
using the partition $h_{S}=\sqrt{\int_{S}\mathcal{E}\left(k_{\perp}\right)d^{2}k_{\perp}}$
where the domain of integration in $k$-space is chosen to correspond
to the injection, dissipation and zonal regions respectively. 

Following Ref. \citealp{berionni:11}, equations (\ref{eq:FE-ZFbis},
\ref{eq:FE-inj}, \ref{eq:FE-dis}) can be averaged over the turbulent
phase, so that time derivatives can be cancelled since the gyrokinetic
simulation reaches a quasi-stationary state as shown in Figure \ref{fig:ZF-NZ-time}.
Coefficients $\lambda_{1}$, $\lambda_{2}$, $\bar{\lambda}$, $\gamma$,
$\nu$ and $\nu_{F}$ are constants in time, and by eliminating the
product $\langle\overline{h}h_{1}h_{2}\rangle$ in the averaged equations,
the two following relations can be obtained:\vspace{-0.5cm}

\begin{eqnarray}
\frac{\langle\mathcal{E}_{1}\rangle}{\langle\overline{\mathcal{E}}\rangle} & = & -\frac{\lambda_{1}\nu_{F}}{\bar{\lambda}\gamma}\,,\label{eq:FEinj/FE-ZF}\\
\frac{\langle\mathcal{E}_{2}\rangle}{\langle\overline{\mathcal{E}}\rangle} & = & \frac{\lambda_{2}\nu_{F}}{\bar{\lambda}\nu}\,.\label{eq:FEdiss/FE-ZF}
\end{eqnarray}

\begin{figure}
\includegraphics[width=1\linewidth]{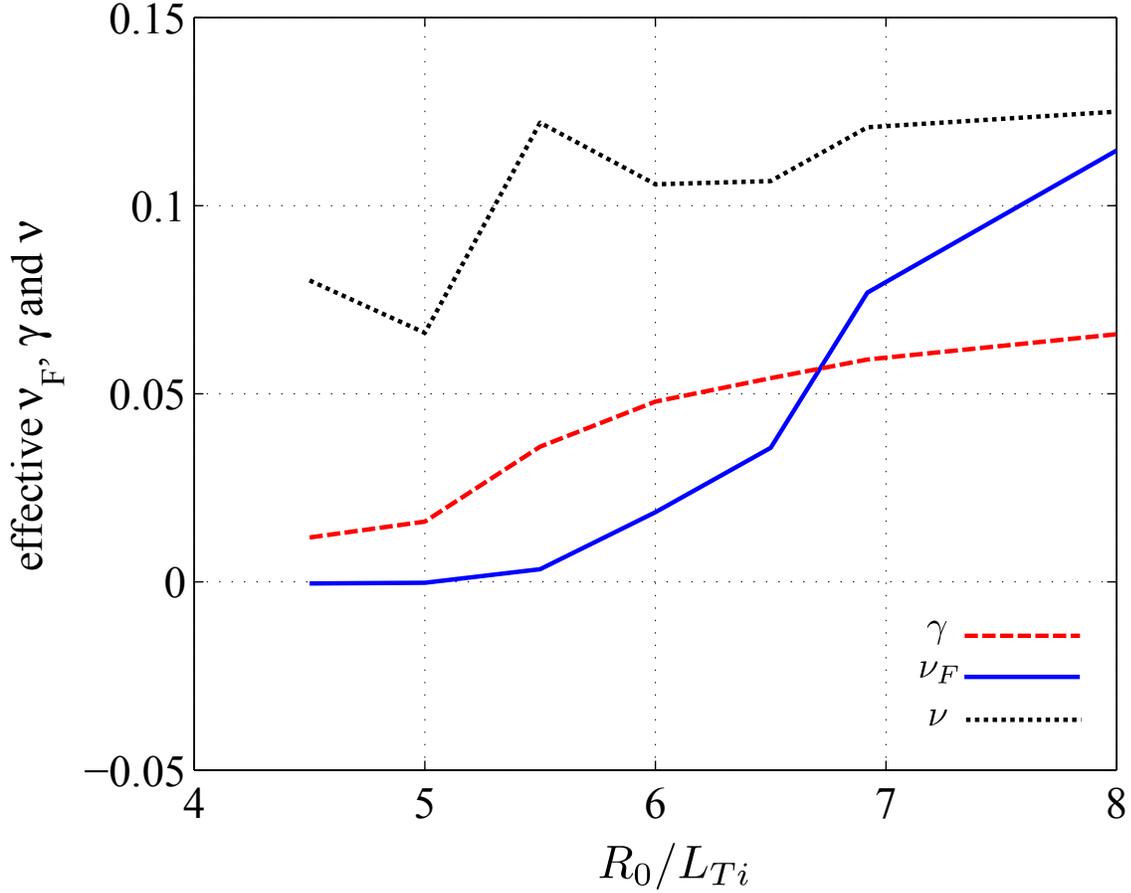}
\caption{\label{fig:cmp-nuF-gamma-nu-LT} Effective Zonal Flows dissipation
($\nu_{F}$ , the solid line), growth rate ($\gamma$ , the dashed
line) and small scale dissipation ($\nu$, the dotted line) as functions
of $R_{0}/L_{Ti}$, with hyperdiffusion GyroLES ($c_{\perp}=0.5$).}
\end{figure}

The parameters $\lambda_{1}$, $\lambda_{2}$ and $\bar{\lambda}$
correspond to geometrical prefactors depending on the choice of the
$k$-space partition and the $C_{{\bf k}}^{{\bf p},{\bf q}}$ that
link $\Phi$ and $h$, and not on the physical parameters. In contrast,
the linear growth rate corresponding to $\gamma$, the small scale
dissipation term $\nu$, and the zonal flow drag $\nu_{F}$ are defined
through $\mathcal{G}_{1,2}$, $\overline{\mathcal{G}}$, $\mathcal{D}_{1,2}$
and $\overline{\mathcal{D}}$, and thus are dependent on the free
energy spectrum itself. Therefore, in principle the net dependence
of these coefficients on physical parameters such as the temperature
gradient can be quite nontrivial. This point is investigated in Figure
\ref{fig:cmp-nuF-gamma-nu-LT}, where the three parameters $\gamma$,
$\nu_{F}$ and $\nu$ are represented as functions of the imposed
$R_{0}/L_{Ti}$.

In Figure \ref{fig:cmp-nuF-gamma-nu-LT}, GyroLES simulations with
various $R_{0}/L_{Ti}$ are considered, with perpendicular hyperdiffusion
amplitude $c_{\perp}=0.5$. The small scale dissipation $\nu$ (dotted
lines) is found approximately constant, except for small values of
the temperature gradient. $\gamma$ and surprisingly $\nu_{F}$ present
a nontrivial dependence with $R_{0}/L_{Ti}$, very similar to the
heat flux structure found in other studies \citep{dimits:00}.

\begin{figure}
\includegraphics[width=0.6\columnwidth]{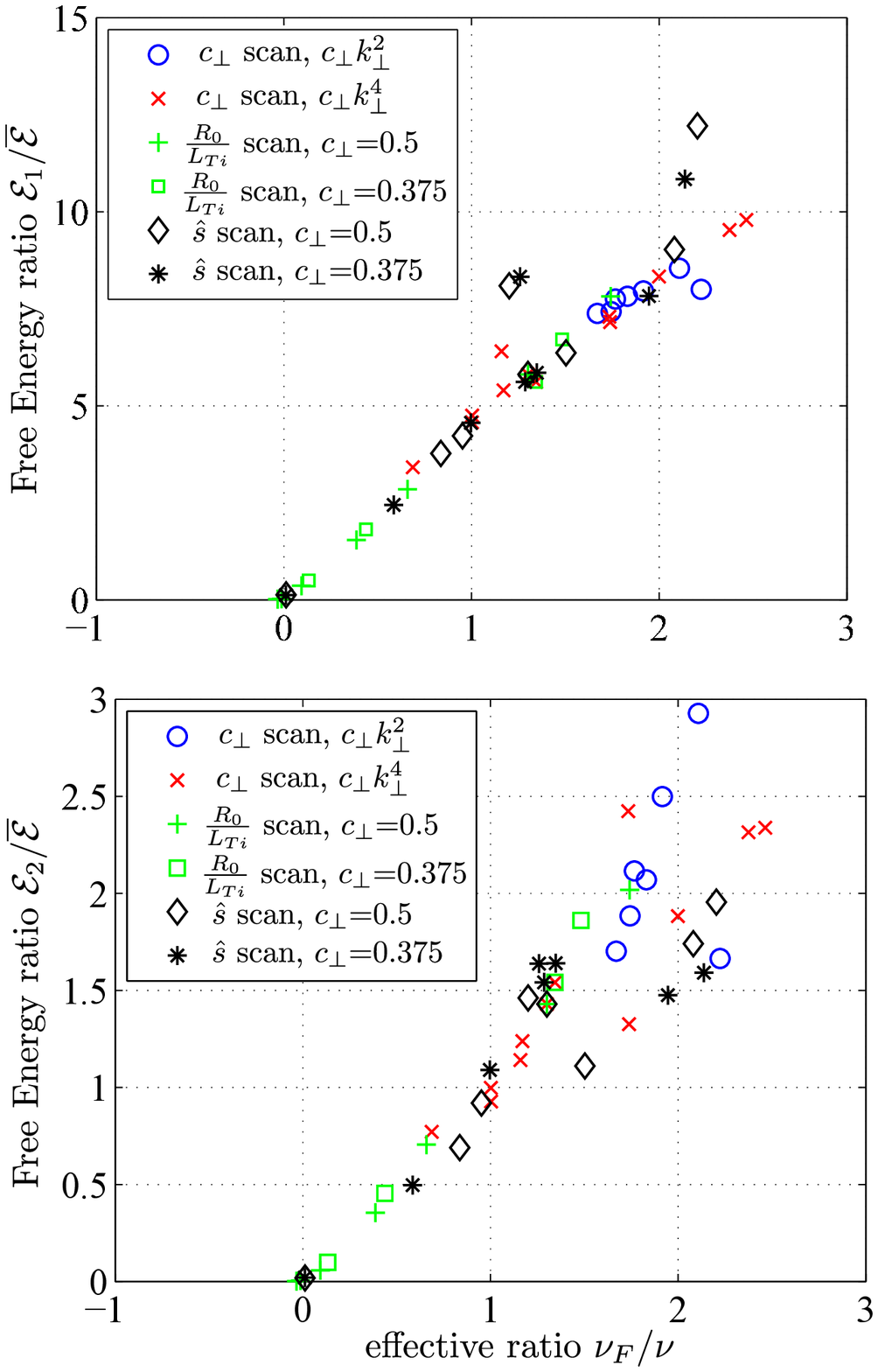}
\caption{\label{fig:E1EZ-nuFgamma-E2EZ-nuFnu} Free energy ratios ($\mathcal{E}_{1}/\overline{\mathcal{E}}$
top, $\mathcal{E}_{2}/\overline{\mathcal{E}}$ bottom) as functions
of the ratio between the effective Zonal Flow dissipation, the effective
growth rate and the effective small scale dissipation (see Eqns \ref{eq:FEinj/FE-ZF}
and \ref{eq:FEdiss/FE-ZF}). Blue circles and red crosses stand respectively
for diffusion and hyperdiffusion GyroLES model scanned along $\ensuremath{c_{\perp}}$,
green plus and squares represent two $R_{0}/L_{Ti}$ scans, with respectively
$\ensuremath{c_{\perp}=0.5}$ and $\ensuremath{c_{\perp}=0.375}$
chosen for hyperdiffusion GyroLES. Black diamonds and stars correspond
to a magnetic shear scan $\hat{s}$ with respectively $c_{\perp}=0.5$
and $c_{\perp}=0.375$.}
\end{figure}

In order to verify Eqns. \ref{eq:FEinj/FE-ZF} and \ref{eq:FEdiss/FE-ZF},
, the averaged free energy ratios $\langle\mathcal{E}_{1}\rangle/\langle\overline{\mathcal{E}}\rangle$
and $\langle\mathcal{E}_{2}\rangle/\langle\overline{\mathcal{E}}\rangle$
are represented as functions of the ratios $\nu_{F}/\nu$ and $\nu_{F}/\gamma$
respectively in Figure \ref{fig:E1EZ-nuFgamma-E2EZ-nuFnu}. Six series
of GyroLES simulations are considered, varying the diffusion or hyperdiffusion
amplitudes, the temperature gradient $R_{0}/L_{Ti}$, as well as the
magnetic shear, for a total of 40 nonlinear GyroLES simulations. The
advantage of using LES for these simulations, (apart from the gain
in speed) is that it provides an easy handle on the small scale dissipation
and allows us to modify $\nu$ independently in order to explore the
parameter space easily.

The curves seems to agree with the theoretically predicted ratios
in Eqns. \ref{eq:FEinj/FE-ZF} and \ref{eq:FEdiss/FE-ZF} . Results
are however found to depart from the theory when the turbulence level
is decreased, especially in the low shear case and for very high perpendicular
dissipation amplitudes. The deviation of the ratio $\mathcal{E}_{1}/\overline{\mathcal{E}}$
from a straight line, is rather small, while it is more pronounced
for $\mathcal{E}_{2}/\overline{\mathcal{E}}$. This could be due to
the reduction of the size of the dissipative range (which affect mainly
\foreignlanguage{english}{$\mathcal{E}_{2}$)} by the use of the GyroLES
description.

We have performed a detailed characterization of the dynamics and
the associated spectral transfer, using the evolution of free energy
in gyrokinetic turbulence with a partition of the $k$-space corresponding
to the scales of free energy injection, free energy dissipation, and
large scale flow structures. Our results show that the free energy
exchange between these components exhibits, any of the well-known
characteristic features of the predator-prey dynamics. We have shown
that the predicted relation between the average amplitudes of these
different components as given in Eqns. \ref{eq:FEinj/FE-ZF} and \ref{eq:FEdiss/FE-ZF}
hold reasonably well in gyrokinetic simulations.

\begin{acknowledgments}
The authors would like to thank F. Jenko and T. Görler for the use
of the GENE code, A. Bañ\'on Navarro and D. Carati for fruitful discussions
and acknowledge that the results in this Letter have been achieved
with the assistance of high performance computing resources on the
HPC-FF systems at Jülich, Germany. This work is supported by the French
\textquotedbl{}Agence nationale de la recherche\textquotedbl{}, contract
ANR JCJC 0403 01 and by the contract of association EURATOM-Belgian
state.
\end{acknowledgments}

\bibliographystyle{apsrev}


\end{document}